\documentclass[]{article}
\usepackage[utf8]{inputenc}
\usepackage{setspace}
\usepackage{epsfig}  
\usepackage{graphicx,epstopdf}  
\usepackage{amsfonts}
\usepackage{amssymb}
\usepackage{amsmath}
\usepackage{subfigure}
\usepackage{comment}
\usepackage{multirow}
\usepackage{psfrag}
\usepackage{soul} 
\usepackage[dvipsnames]{xcolor}
\usepackage{fullpage}

\begin{document}

\title{Throughput Analysis in High Density WLANs} 

\author{Boris Bellalta \\ Dept. of Information and Communication Technologies \\ Universitat Pompeu Fabra}

\date{}

\maketitle

\begin{abstract}
This letter proposes a simple but accurate approximation to analytically model both the inter-Wireless Local Area Networks (WLANs) interactions and the negative effect of collisions in the performance of high density WLAN scenarios. Inter-WLANs interactions are characterized using a continuous time Markov chain (CTMC) model where states represent the set of active WLANs at a given time. Then, the effect of collisions is considered by analyzing the local dynamics between contending WLANs at every state of the CTMC. Simulation results confirm the accuracy of the presented approach.
\vspace{0.5cm}  
  
\end{abstract}

\onehalfspacing


\section{Introduction}

Analyzing scenarios with multiple IEEE 802.11 Wireless Local Area Networks (WLANs) operating in the same channel when not all of them are within each other's carrier sense range is challenging. That is because in this case, the operation of the WLANs, which is governed by the Carrier Sense Multiple Access with Collision Avoidance (CSMA/CA) protocol, is not fully coupled since not all of them observe the same events.

To characterize the interactions between WLANs, continuous time Markov chain (CTMC) models have recently\footnote{The use of CTMC models to analyze CSMA/CA networks is not new. They have been extensively used in the last years to capture the interactions between nodes in multi-hop networks \cite{boorstyn1987throughput,laufer2013capacity,liew2010back}.} been proposed \cite{baid2015understanding,bellalta2016interactions,michaloliakos2016performance}\footnote{ \cite{baid2015understanding,michaloliakos2016performance} focus on the \textit{maximal independent sets} approach presented in \cite{liew2010back} to compute the throughput of each WLAN, which only considers the maximal states of the CTMC model. Maximal states are those in which the maximum possible number of nodes are transmitting simultaneously.}. A CTMC model describes all possible combinations of active WLANs (states) as well as the transition rates between states. Thus, CTMC models allow, building upon the assumption of exponentially distributed backoff and packet transmission durations, to compute the long-term fraction of time the system remains at each state.

CTMC models are able to capture the interactions between neighboring WLANs but their accuracy decreases as collisions become increasingly significant, which has raised many concerns about their use in performance evaluation of next-generation, high density WLANs. The reason for this is the need in CTMC models to assume continuous backoff timers, which, as a consequence, results in negligible collision probability (i.e., the probability that the backoff timer of two or more nodes expire simultaneously). Hence, the use of CTMC models to characterize the performance of WLANs using a slotted backoff counter such as that defined in the IEEE 802.11 standard is only accurate in the case of small number of contenders or large backoff windows.

This letter provides a simple but accurate approximation to analytically model both the inter-WLANs interactions and the negative effect of collisions in scenarios with multiple adjacent WLANs. First, inter-WLANs interactions are characterized using a WLAN-centric CTMC model that allows us to derive the fraction of time each WLAN is active. Then, to compute the corresponding fraction of the WLAN's active time that corresponds to successful transmissions, we analyze the local interactions at the node level between contending WLANs at every state of the CTMC using the well-known Bianchi IEEE 802.11 model\footnote{Note that other equivalent IEEE 802.11 analytical models could be used instead of Bianchi's model. We have selected Bianchi's model due to its accuracy and well-known conditions under which it can be applied.} \cite{bianchi2000performance}. Using Bianchi's model we estimate the collision probability, which is then normalized to cope with the different temporal network dynamics of using a slotted backoff timer instead of a continuous one. Simulation results confirm the accuracy of the presented approach.


\section{Modelling Inter-WLANs interactions}
\label{Sec:interWLAN}

The analysis presented in this section is based in \cite{bellalta2016interactions}, where the considerations required to derive a WLAN-centric CTMC model are discussed in detail. The work in \cite{bellalta2016interactions} also evaluates the computational advantage of the WLAN-centric CTMC approach compared to the node-centric CTMC one. 

\subsection{Multiple WLANs scenario}

A scenario consisting of multiple WLANs deployed in a certain area is considered. WLAN~$i$ consists of $N_i$ nodes, the access point (AP) and the $N_{i}-1$ associated stations (STAs). Following the CSMA/CA operation, whenever a transmission from another WLAN is detected, the backoff of WLAN~$i$ is paused until the channel is detected free again, at which point the countdown is resumed. The expected active backoff duration for WLAN~$i$, i.e., excluding the fraction of time the backoff is paused, is $1/\lambda_i$. At every transmission, WLAN~$i$ sends $L$ bits and occupies the channel for an expected duration equal to $E[T]=1/\mu$. A full-buffer traffic model is considered for all nodes.

For simplicity, we assume that when coverage ranges from two WLANs overlap, the coverage ranges of all nodes in the WLANs involved also overlap. Such an assumption covers scenarios where there are multiple side-by-side WLANs, such as in an apartment building, where all STAs belonging to a WLAN are close to the AP. 

\subsection{CTMC model}

Let us define a feasible system state as the subset of WLANs that can transmit simultaneously without interference, i.e., WLANs with non-overlapping coverage ranges. Let $\Omega$ be the collection of all feasible system states. Then, the transition rates between two system states $s', s \in \Omega$ are
\begin{equation}
q(s',s)=
\begin{cases}
\lambda_{i} & \text{ if } s=s' \cup \{i\} \in \Omega,\\
\mu & \text{ if } s=s' \setminus \{i\},\\
0 & \text{ otherwise}.  \nonumber
\end{cases}  
\end{equation}

Let $S_t \in \Omega$ be the system state at time $t$. Assuming exponentially distributed backoff and transmission durations, $(S_t)_{t\geq 0}$ is a continuous-time Markov process on the state space $\Omega$. This Markov process is aperiodic, irreducible and thus positive recurrent, since the state space $\Omega$ is finite. Hence, it has a stationary distribution, which we denote by $\{\pi_s \}_{s \in \Omega}$. It follows from classical Markov chains results that $\pi_{s}$ is equal to the long-run fraction of time the system is in state $s \in \Omega$.

The process $(S_t)_{t\geq 0}$ has been proven to be a time-reversible Markov process in~\cite{kelly1979reversibility}. In particular, detailed balance applies and the stationary distribution $\{\pi_s \}_{s \in \Omega}$ of the process $(S_t)_{t \geq 0}$ can be expressed in product form. The detailed balance relationship for two adjacent system states, $s$ and $s \cup \{i \}$, can be written as follows:
\begin{align}
\frac{\pi_{s\cup \{i\}}}{\pi_{s}} = \frac{\lambda_{i}}{\mu}=\theta_{i}.  \nonumber
\end{align}

This relationship implies that for any $s \in \Omega$
\begin{align}
\pi_{s} = \pi_{\emptyset} \cdot \prod_{i \in s} \theta_{i}, \nonumber
\end{align}
where $\emptyset$ denotes the system state where none of the WLANs is transmitting. The last equality, together with the normalizing condition $\sum_{s \in \Omega} \pi_s = 1$, yields
\begin{align}
\pi_{\emptyset} = \frac{1}{\sum_{z \in \Omega} \prod_{j \in z} \theta_{j}},  \nonumber
\end{align}
and
\begin{align}
\pi_{s} = \frac{\prod_{i \in s} \theta_{i}}{\sum_{z \in \Omega} \prod_{j \in z} \theta_{j}}, \quad s \in \Omega.  \nonumber
\end{align}


\section{Throughput computation}
\label{Sec:Throughput}

The throughput of WLAN~$i$ is given by
\begin{align}\label{Eq:Throughput}
    x_{i} &=  \left(\sum_{s \in \Omega\,:\, i \in s}{\pi_s (1-\gamma_{i|s'\rightarrow s})}\right)\mu L,
\end{align}
where $1-\gamma_{i|s'\rightarrow s}$ is the probability that the data transmitted by WLAN~$i$ in state $s$ (moving from a predecessor state $s'$ in which WLAN~$i$ is not active) is successfully received, and $L$ is the packet size. Since we do not consider transmission errors, the value of $\gamma_{i|s'\rightarrow s}$ is caused only by packet collisions. Following the IEEE 802.11 slotted backoff, a collision occurs when the backoff counter of two or more contending nodes reaches zero at the same time. We consider that all packets involved in a collision are lost.

\subsection{Conditional throughput from state $s'$}

Let us denote by $\mathcal{K}_{i|s'}$ the set of contending WLANs (i.e., with overlapping coverage ranges) with WLAN~$i$ to access the channel at state $s'$, and $k_{i|s'}=\sum_{j \in \mathcal{K}_{i|s'}}{N_{j}}$ the total number of nodes in $\mathcal{K}_{i|s'}$. 

The achievable throughput by WLAN~$i$ when it is contending with the WLANs in $\mathcal{K}_{i|s'}$ is 
\begin{align}\label{Eq:y}
    y_{i|s'}=\frac{d_{i|s'}L}{a_{i|s'}T_e+b_{i|s'}E[T]+c_{i|s'}E[T_c]},
\end{align}
where $a_{i|s'}=(1-\tau_{i|s'})^{k_{i|s'}+N_{i}}$ is the probability that a backoff slot remains empty, $b_{i|s'}=(k_{i|s'}+N_{i})\tau_{i|s'}(1-\tau_{i|s'})^{k_{i|s'}+N_{i}-1}$ is the probability that a slot results in a successful transmission, $c_{i|s'}=1-a_{i|s'}-b_{i|s'}$ is the probability that a slot results in a collision, and $d_{i|s'}=N_{i}\tau_{i|s'}(1-\tau_{i|s'})^{k_{i|s'}+N_{i}-1}$ is the probability that a slot results in a successful transmission from WLAN~$i$. $T_e$ and $E[T_c]$ are the duration of an empty slot and a collision, respectively.

To compute $\tau_{i|s'}$, the probability to transmit in a given slot, we solve the following system of equations using a fixed-point method:
\begin{align}
\left \{\begin{array}{l}
    E[B_{i|s'}]=\frac{1 - p_{i|s'} - p_{i|s'}(2p_{i|s'})^{m}}{(1-2p_{i|s'})}\frac{CW_{\min}}{2}-\frac{1}{2} \nonumber \\
    p_{i|s'}=1-(1-\tau_{i|s'})^{k_{i|s'}+N_{i}-1} \nonumber\\
    \tau_{i|s'} = \frac{1}{E[B_{i|s'}]+1} \nonumber
    \end{array}\right.
\end{align}
where $E[B_{i|s'}]$ is the expected backoff duration in slots, $p_{i|s'}$ is the collision probability, and $m=\log_2\left(\frac{\text{CW}_{\max}}{\text{CW}_{\min}}\right)$ is the relationship between the maximum and minimum backoff contention window as defined in the IEEE 802.11 standard.

Then, $\gamma_{i|s'\rightarrow s}$ is computed so as to result in the same throughput as (\ref{Eq:y}) when (\ref{Eq:Throughput}) is used to compute the throughput of $\mathcal{K}_{i|s'}+1$ WLANs contending for the channel, i.e.,
\begin{align}
    \mu L\left(\frac{\theta_{i}}{1+\theta_i+\sum_{j \in \mathcal{K}_{i|s'}}{\theta_{j}}}\right) (1-\gamma_{i|s'\rightarrow s}) = y_{i|s'}, \nonumber
\end{align}
from where the value of $\gamma_{i|s'\rightarrow s}$ is obtained:
\begin{align}\label{Eq:gamma}
    \gamma_{i|s'\rightarrow s}=1-\frac{y_{i,s'}}{\mu L\left(\frac{\theta_{i}}{1+\theta_i+\sum_{{j} \in \mathcal{K}_{i|s'}}{\theta_{j}}}\right)}, 
\end{align}
with $\theta_i=\frac{\lambda_i}{\mu}$ and $\lambda_i=N_i\frac{2}{CW_{\min}-1}\frac{1}{T_e}$.

\subsection{Relationship between $\gamma_{i|s'\rightarrow s}$ and $p_{i|s'}$}

\begin{figure}[t!!!!!!!!]
\centering
\epsfig{file=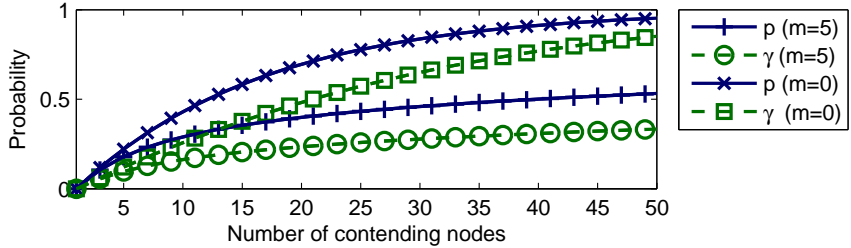,width=0.7\columnwidth,angle=0}
\caption{Relationship between $\gamma_{i|s'\rightarrow s}$ and $p_{i|s'}$, for $\text{CW}_{\min}=32$. }
\label{Fig:Fig.0}
\end{figure}

Intuitively, we could have considered that $\gamma_{i|s'\rightarrow s} = p_{i|s'}$ since $p_{i|s'}$ is the fraction of transmissions colliding in WLAN~$i$. However, such an approximation for $\gamma_{i|s'\rightarrow s}$ results in very conservative predictions. The reason for this is that the rate at which packets are transmitted to the channel in the continuous-time backoff case is lower than when the IEEE 802.11 backoff is used, because collisions represent multiple concurrent transmissions. Therefore, removing the same fraction of transmissions in the former case as in the later is not accurate. Then, we need to normalize the value of $\gamma_{i|s'\rightarrow s}$ to the actual rate of packet transmissions in the continuous time backoff case, hence obtaining that $\gamma_{i|s'\rightarrow s} \leq p_{i|s'}$ (see Fig. \ref{Fig:Fig.0}).

\subsection{A note on ``dominant states''-only consideration}

Considering typical IEEE 802.11 $\text{CW}_{\min}$ and $E[T]$ values, WLANs remain a high fraction of time (e.g., $\geq$ 95 \%) in a subset of states of the CTMC to which we refer as \textit{dominant states}.  Therefore, the computational cost to solve (\ref{Eq:Throughput}) can be reduced by considering only these dominant states and its predecessors without a significant effect on the accuracy of the obtained results. \textit{Dominant states} correspond to the \textit{maximal independent sets} defined in \cite{liew2010back} when the CSMA/CA operational parameters (e.g., channel width, Clear Channel Assessment (CCA), transmit power, etc.) are static. Otherwise, when these parameters are set depending on the system state, as it is shown in \cite{faridi2015analysis} for the case of Dynamic Channel Bonding, we obtain dominant states that are not maximal, thus it requires exploration of the full state space $\Omega$ to determine them.


\section{Examples}
\label{Sec:Examples}

\begin{figure}[th!!!!!!!!]
\centering
\includegraphics[width=0.7\columnwidth]{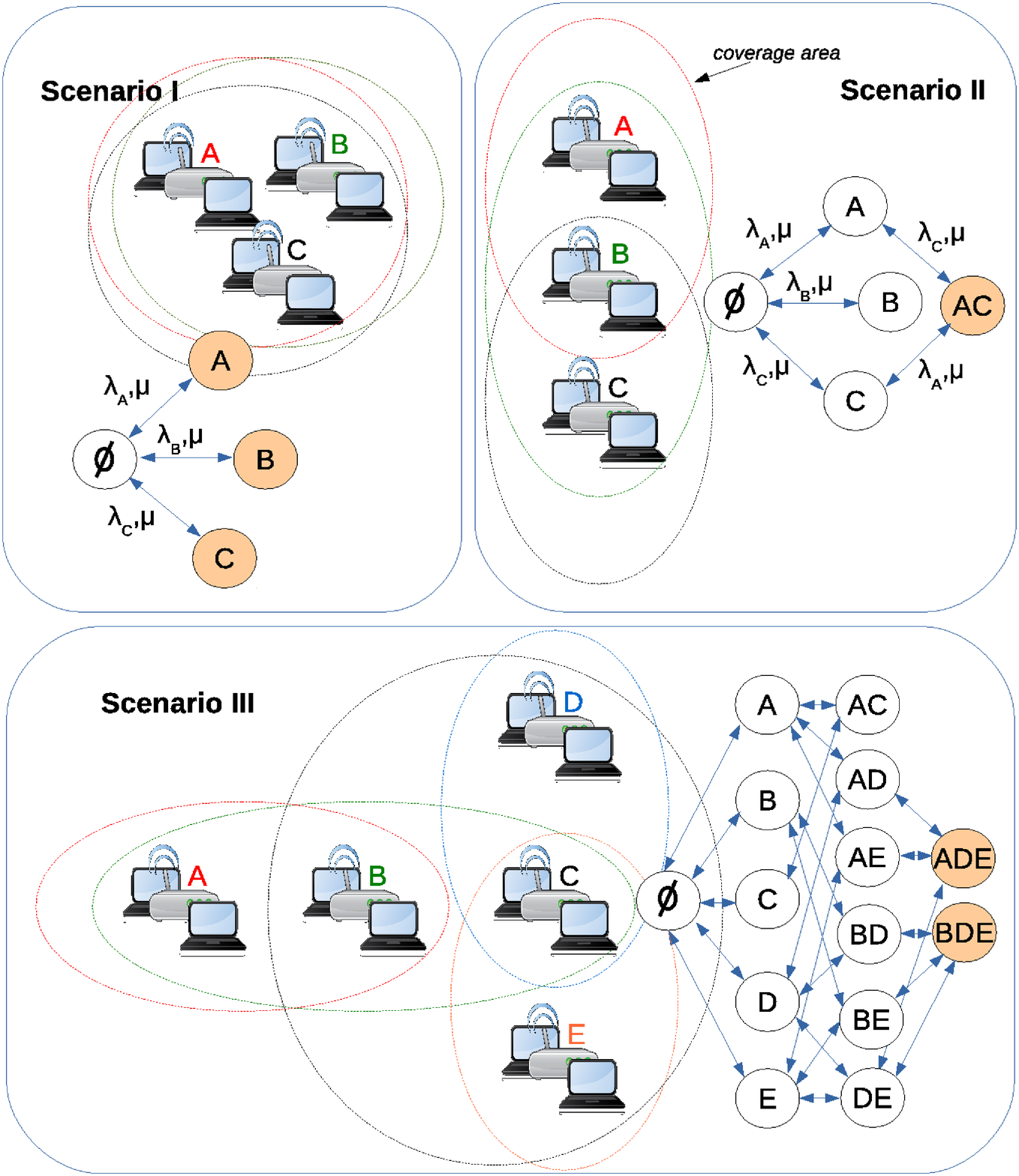}
\caption{The three scenarios (Scenario I, II and III) used in Sec. \ref{Sec:Examples}. We consider all WLANs have the same number of nodes ($N$), i.e., 1 AP and $N-1$ STAs, and share the same values of $\lambda$ and $\mu$. In each CTMC, the states in orange represent the maximal states. Forward ($\lambda$) and backward ($\mu$) transition rates are represented over single double-headed arrows for Scenario I and II. For Scenario III they are omitted for space reasons.}
\label{Fig:Networks}
\end{figure}

To show the accuracy of the proposed method when computing the throughput of each WLAN in the system, we compare the throughput obtained by the proposed approach in Sec. \ref{Sec:Throughput}, with simulation results in the three scenarios depicted in Fig. \ref{Fig:Networks}. For each scenario  we indicate the position and coverage area of each WLAN, and include the CTMC model that represents the feasible system states.

The simulator was developed in C using the Component Oriented Simulation Toolkit libraries \cite{chen2002reusing}. It accurately reproduces the described scenarios and the operation of each WLAN. The parameters used in this section correspond to the IEEE 802.11ac amendment, with 40 MHz channels, 64-QAM modulation and 3/4 coding rate. Packet aggregation is considered, and each transmitted Aggregated MAC protocol data unit (A-MPDU) contains 64 packets of length $12000$ bits, resulting in a packet size of $L=768$ Kbits. Under these conditions, the expected duration of a successful transmission is given by $E[T]=6.63$ ms (for further details on how $E[T]$ is calculated, please refer to \cite{bellalta2016interactions}). Since the RTS/CTS mechanism is not used, we consider that $E[T_c]=E[T]$. Finally, unless otherwise specified, $\text{CW}_{\min}=32$ and $m=5$.

\begin{figure}[th!!!!!!!!]
\centering
\subfigure[Throughput achieved by each WLAN in Scenario I.]{\epsfig{file=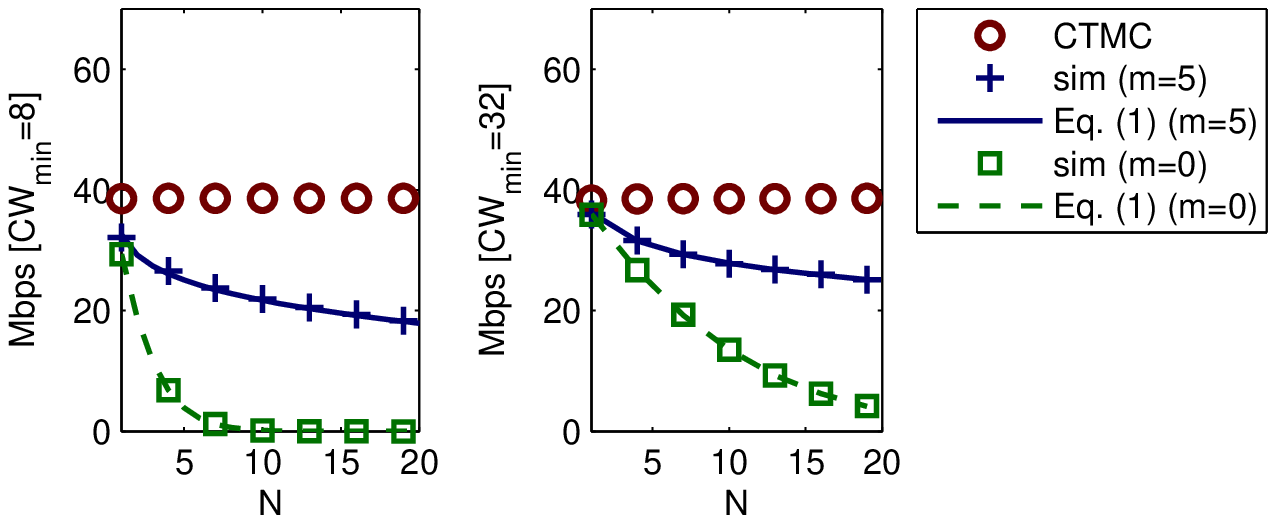,width=0.7\columnwidth,angle=0}\label{Fig:Fig.1}}
\quad
\subfigure[Throughput achieved by each WLAN in Scenario II.]{\epsfig{file=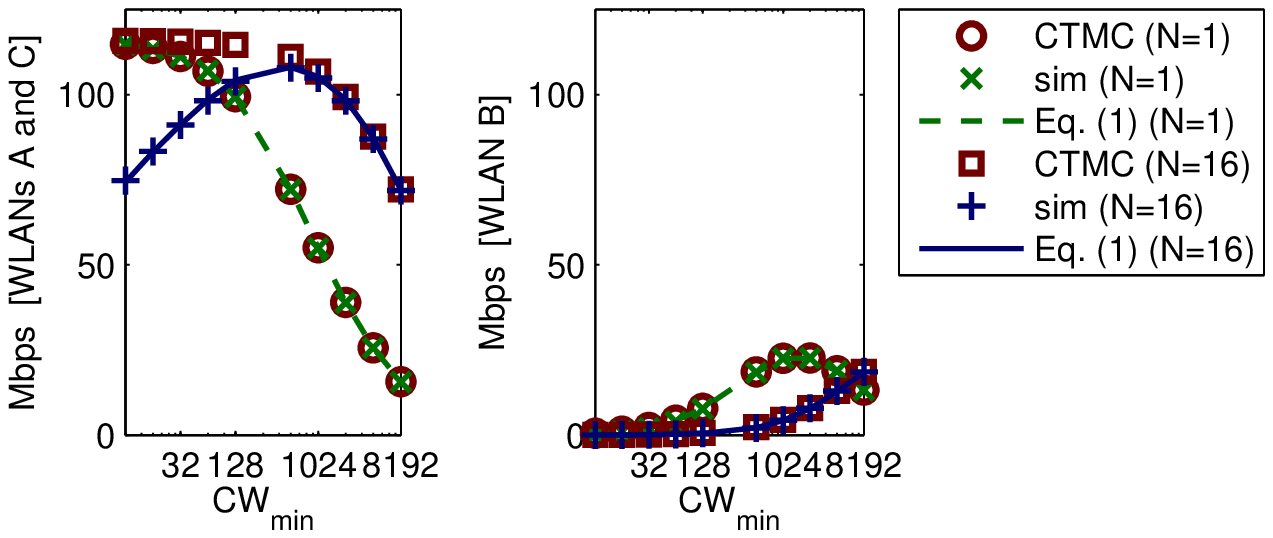,width=0.7\columnwidth,angle=0}\label{Fig:Fig.2}}
\quad
\subfigure[Throughput achieved by each WLAN in Scenario III.]{\epsfig{file=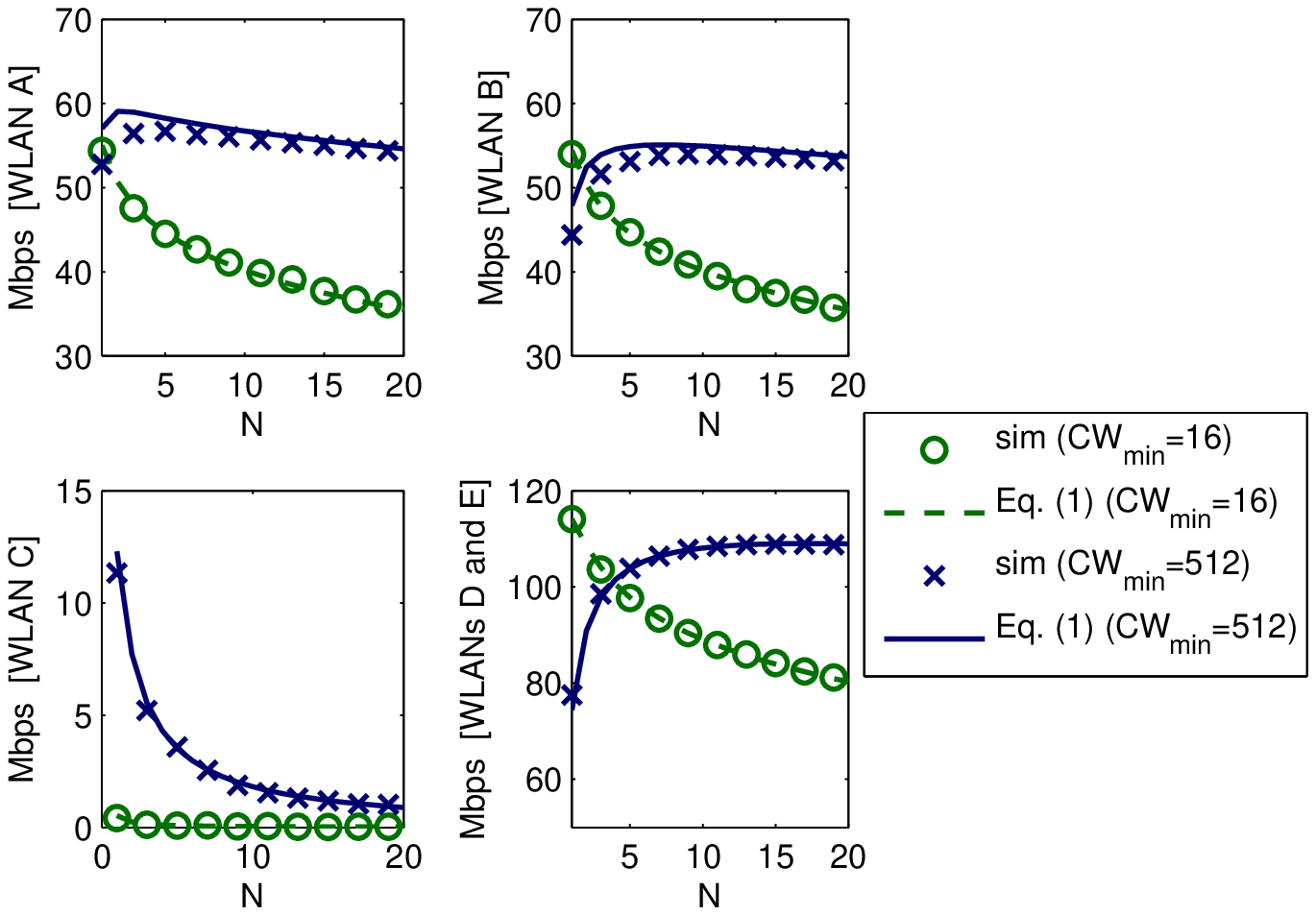,width=0.7\columnwidth,angle=0}\label{Fig:Fig.4}}
\caption{Throughput achieved in each scenario. In the legend, by 'CTMC' and 'sim' we refer to the throughput obtained using (\ref{Eq:Throughput}) with $\gamma_{i|s'\rightarrow s}=0$ and by simulation, respectively.}
\label{Fig:Results}
\end{figure}

The first scenario considers three WLANs with fully overlapping coverage ranges. From (\ref{Eq:Throughput}), the throughput of WLANs A, B and C is given by
\begin{align}
    x_A=x_B=x_C = \mu L\frac{\theta(1-\gamma_{C|\emptyset\rightarrow C})}{1+3\theta}, \nonumber
\end{align}
where $\theta_A=\theta_B=\theta_A=\theta$.

Fig. \ref{Fig:Fig.1} shows a very good agreement between the throughput values provided by the analysis and simulation results. Notice the difference between the throughput estimated by the CTMC model (i.e., using (\ref{Eq:Throughput}) with $\gamma_{i|s'\rightarrow s}=0$), which is independent of $m$ as it assumes negligible collision probability, with the throughput values obtained from the simulator and the proposed approximation (i.e., using (\ref{Eq:Throughput}) with $\gamma_{i|s'\rightarrow s}$ from (\ref{Eq:gamma})).

The second scenario consists of three WLANs placed in a line, with WLAN B in the middle of WLANs A and C. In this scenario, the throughput achieved by WLANs A, B and C is given by
\begin{align}
    x_A=x_C&=\mu L\frac{\theta(1-\gamma_{\text{C}|\emptyset\rightarrow \text{C}})+\theta^2(1-\gamma_{\text{C}|\text{A}\rightarrow \text{AC}})}{1+3\theta+\theta^2}, \text{and} \nonumber \\
    x_B &= \mu L\frac{\theta(1-\gamma_{\text{B}|\emptyset\rightarrow \text{B}})}{1+3\theta+\theta^2}. \nonumber
\end{align}

Fig. \ref{Fig:Fig.2} shows the throughput obtained by WLANs A, B and C when the value of $\text{CW}_{\min}$ increases from $4$ to $8192$. It can be observed that the proposed approximation is very accurate for the two values of $N$. For $N=1$, the CTMC model shows good accuracy as well since collisions can be neglected. However, for $N=16$ and low $\text{CW}_{\min}$ values, the negative effect of collisions in the throughput of WLANs A and C, which the CTMC model fails to capture, is evident. 

Scenario III extends Scenario II by adding WLANs D and E as shown in Fig. \ref{Fig:Networks}. Note that WLANs D and E coverage ranges do not overlap, and therefore severely affect the transmission opportunities of WLAN C. In this case, WLAN C chances to access the channel are severely reduced due to the activity of WLANs D and E. Consequently, WLAN B is positively affected as it only (or almost only) contends with WLAN A. In the absence of WLANs D and E, on the contrary, WLANs C and A reduce WLAN B transmission opportunities as shown in Scenario II.

The throughput of each WLAN in Scenario III is given by (writing only the most significant states for each WLAN)
\begin{align}
    x_A & =\mu L \left(\frac{\ldots+\theta^3(1-\gamma_{\text{A}|\text{DE}\rightarrow \text{ADE}})}{1+5\theta+6\theta^2+2\theta^3}\right), \nonumber \\    
    x_B & =\mu L \left(\frac{\ldots+\theta^3(1-\gamma_{\text{B}|\text{DE}\rightarrow \text{BDE}})}{1+5\theta+6\theta^2+2\theta^3}\right), \nonumber \\
    x_C & =\mu L\left(\frac{\ldots+\theta^2(1-\gamma_{\text{C}|\text{A}\rightarrow \text{AC})}}{1+5\theta+6\theta^2+2\theta^3}\right), \text{~and}\nonumber \\
    x_D & =x_E= \nonumber \\ &= \mu L\left(\frac{\ldots+\theta^3(1-\gamma_{\text{E}|\text{BD}\rightarrow \text{BDE})}+\theta^3(1-\gamma_{\text{E}|\text{AD}\rightarrow \text{ADE}})}{1+5\theta+6\theta^2+2\theta^3}\right). \nonumber 
\end{align}

Fig. \ref{Fig:Fig.4} confirms the accuracy of the presented approach in Scenario III. The accuracy of the approximation is reduced for $\text{CW}_{\min}=512$ compared to $\text{CW}_{\min}=16$. Indeed, large $\text{CW}_{\min}$ values affect the backoff dynamics of each contender when the system reaches state $s'$, not abiding to the stationary behavior assumed in (\ref{Eq:y}). 


\section{Conclusions}
\label{Sec:Conclusions}

This letter presents a simple approach to approximate the achievable throughput of a group of WLANs operating in a given area, capturing both their mutual interactions and the negative effect of collisions in the system performance. It is based on decoupling the 'macro' interactions caused by backoff pauses when the channel is detected busy, from the 'micro' interactions caused by packet collisions resulting from two or more nodes' backoff timers expiring at the same time. The accuracy of the results confirms that such 'macro' and 'micro' interactions have an important effect in high density WLAN deployments, and they can be considered as independent in most cases.

While in this letter we have evaluated the presented approach in several basic scenarios, there are many other aspects to consider in the future to determine under which conditions the presented approach is accurate, providing also further insight into the operation of high density WLANs. For example, future research extensions include: \emph{i)} considering the presence of partially overlapping coverage ranges, encompassing the effect of hidden and exposed terminals, \emph{ii)} taking account of the capture effect, and \emph{iii)} incorporating miscellaneous traffic conditions. The proposed approach must also be validated for WLANs employing advanced features, such as static and dynamic channel bonding, CCA adaptation and multiuser communications \cite{bellalta2016ieee}.


\bibliographystyle{unsrt}
\bibliography{Bib}

\end{document}